\documentclass[aps,prd,twocolumn,groupedaddress,showpacs,nofootinbib,amssymb]{revtex4}

\begin{document}

\title{Majorana Tower and Cellular Automaton Interpretation of Quantum Mechanics down to Planck Scales}
\author{Fabrizio Tamburini} 
\email{fabrizio.tamburini@gmail.com}
\affiliation{o.l.f. IPSEOA Barbarigo, Venice, Italy}
\affiliation{ZKM -- Zentrum f\"ur Kunst und Medientechnologie, Lorentzstr. 19, D-76135, Karlsruhe, Germany.}

\author{Ignazio Licata}
\email{ignaziolicata3@gmail.com}
\affiliation{Institute for Scientific Methodology (ISEM) Palermo Italy}
\affiliation{School of Advanced International Studies on Theoretical and Nonlinear Methodologies of Physics, Bari, I-70124, Italy}
\affiliation{International Institute for Applicable Mathematics and Information Sciences (IIAMIS), B.M. Birla Science Centre, Adarsh Nagar, Hyderabad -- 500 463, India}

\begin{abstract} 
A deterministic reformulation of quantum mechanics can bypass the usual philosophical interpretations of probability and stochasticity that are found in the literature. This can be obtained with the ontological formulation of quantum mechanics, obtained by writing the Hamiltonian of a quantum system in a way to render it mathematically equivalent to a deterministic system. Such deterministic models are thought to consist of elementary cells - cellular automata - inside which the quantities describing the dynamics oscillate in periodic orbits, extending and replacing the quantum-mechanical classical language based on harmonic oscillators.
Here we show that the structure of the cellular automata sets find a clear physical interpretation with the infinite-components equation published by Majorana in 1932, also known as the Majorana Tower: the cellular automata are elementary building blocks generated by the Poincar\'e group of spacetime transformations with positive-defined energy down to the elementary building blocks of the fabric of spacetime. 
Interestingly, the mathematical approach here considered presents close relationships with those used for the distribution of prime numbers in the P\'olya-Hilbert conjecture for the Riemann Hypothesis.
\end {abstract}
\maketitle

\section{Introduction}
The never ended debate whether quantum mechanics (QM) is either stochastic or fully deterministic in its intrinsic nature takes its birth from the historical Bohr-Einstein debate in 1935 \cite{marage}; there may also exist a more subtle level of the discrete deterministic type that gives the stochastic behavior as output. Then, in 1964, Bell's theorem \cite{bell,bell2,bell3} put an end to many ideal scenarios where hidden variables were the only engine for randomness, ideally favoring models where non locality arises \cite{scarani,bohm}.

The ontological quantum mechanics (OQM) here considered \cite{thooft,licata} is a reformulation of QM slightly different from the classical probabilistic formulations \cite{diracbook,Hossen2011,Hossen2020}. In OQM the Hamiltonian of a quantum system is rendered mathematically equivalent to that of a deterministic system with a novel mathematical language that describes structures evolving deterministically. This language can be used to describe the evolution of both a quantum and a classical system. This is a step forward the debate started with the discussions between Bohr and Einstein, where Einstein did never accept the intrinsic stochasticity present in the language of quanta, whilst Bohr did. 
Differently from the well-known hidden variable theories falsified by both the theory and the experiments on Bell's inequalities \cite{zeilinger2,zeilinger3}, this new approach is robust with respect to the Einstein-Podolski-Rosen paradox and the problem of hidden variables \cite{epr,thooft2}.

The quantum mechanical ``ontological'' states are represented by sets of orthonormal unit vectors in the Hilbert space of support that can be either finite or infinite-dimensional. By definition, a system is ontological if it evolves in time into other ontological states, with no difference between classical quantum and deterministic state.
Locally ontological and deterministic systems can be constructed that can feature quantum mechanical properties including entanglement and the violation of Bell's inequalities.
The classical dynamical system varies with time scales much smaller than the time related to the energy exchanged in any interaction there considered, $\Delta t \ll 1/\Delta E_{int}$. The system is deterministic if from ontological states it evolves into other ontological states and any state either classical or quantum is identified by a ket vector $| n \rangle$. Of course one can have systems with continuous dynamics or with cyclic one. As usual, if the evolutional time step is discrete, then the Hamiltonian is periodic in its eigenvalues, introducing the concept of ``beable'', a vector state, as proposed by Bell to replace the traditional term observable of QM that might imply the interaction with an observing device or measurement issue and that of ``changeable'', ``superimposable'' and non-local phenomena, associated with cellular automata (CA) in Hilbert spaces \cite{thooftbook}.

\section{Deterministic systems}

Here we analyze two main classes of deterministic systems leading to an ontological deterministic representation of quantum mechanics. The continuous  systems have a set of equations describing a continuous dynamics whose QM-type indeterminism is due to a discretization in time or, equivalently, to a tassellation of the phase space, that can go down the Planck scales, as occurs in the search of the distribution of prime numbers with the Hilbert-P\'olya approach, where are searched Hermitian Hamiltonians with eigenvalues that describe the distribution of the zeros of the Riemann's zeta function \cite{schumi}.
The second class is instead made of periodic models with $SU(2)$ structure described by the infinite-components Majorana equation \cite{Majorana:NC:1932}.

\subsection{Continuous deterministic systems}
The deterministic nature of a given physical system is revealed through the analysis of the eigenvalue spectrum of its Hamiltonian that can be written as
\begin{equation} 
H=T(\vec p\,)+V(\vec x\,)+\vec A(\vec x\,)\cdot p\ , 
\end{equation}
where $\vec x$ and $\vec p$ are the usual coordinates and momenta, for which the usual $[x_i,\,p_j]=i\delta_{ij}$ holds. The kinetic term $T(\vec p) \sim 1/2 {\vec p}^{~2}$ and the classical potential $V(\vec x)$, responsible for the change of the geometry of the trajectories (and of spacetime, see \cite{riemannft}), represent the usual constituents of a standard Hamiltonian that one can find in both continuous and quantum systems, where interference patterns are present. A route to chaos and randomness from a continuous deterministic system is clearly provided if one considers as simple example the so-called magnetic term of the Hamiltonian only, $\vec A(\vec x\,)\cdot p$; this alone can describe a route to chaos when a Heisenberg-like texture is introduced in the phase space of a system with Hamiltonian $H=\vec A(\vec x\,)\cdot p$. In the simplest case, $\vec A(\vec x\,) = x$ when one assumes a lattice geometry for the time coordinate the Hamiltonian eigenvalues can also become periodic, with similarities to Berry-Keating and Connes \cite{berry1,berry2,connes} semiclassical dynamics based on the class of $H=xp$ Hamiltonians that have been used in the attempts to solve the Riemann hypothesis from the Hilbert-P\'olya approach. The Riemann Hypothesis is true if there exists a Hermitian or unitary operator whose eigenvalues distribute like the zeros of Riemann's $\zeta(z)$. In this way, space, time, and often also momentum, can be considered discrete. As described in the literature of prime number distribution, magnetic-term dominated Hamiltonians cannot always be Hermitian, showing the properties of PT-symmetric quantum systems \cite{strumia,bender} unless after some modifications and ad-hoc assumptions to the phase space which becomes rigged \cite{delamadrid}. Following the idea by Hilbert and P\'olya, Hermitian Hamiltonians of this type of dynamical systems can describe the distribution of the zeros of Riemann's zeta function and thus of primes \cite{riemannft}, connecting two apparently distant worlds: the fabric of spacetime and the fabric of prime numbers.

The limit to the lattice size, both for continuous and periodic dynamical systems finds its roots down to the Planck scale where the problems of undefined time coordinate below the Planck time $\tau_P$ or their equivalence with both spatial and temporal coordinates are instead described by an indetermination relationship directly derived from Einstein's equations for the scalar proper energy $E$ - averaged over a proper volume $L^3$ - and the corresponding interval of time $\tau$ \cite{erepr}. 
In this case the lattice is directly provided by the fluctuations in the fabric of spacetime. What is important to point out is that Einstein's equations and deterministic continuous dynamical systems can hold down to the Planck scale, with a dynamics recalling that in a Minkowski spacetime with a lattice structure.
The lattice-like structure is given by the indetermination relationship between the proper energy $E$ averaged over a given proper volume $L^3$ in General Relativity (GR),
\begin{equation}
\left< E \right> = \bar E\sim \frac{g^2}{L}R_{(4)}= L \left( \Delta \left(\frac{\Delta g}{g}\right) + \left(\frac{\Delta g}{g}\right)^2\right),
\end{equation}
where $g$ is the metric tensor, $\Delta g$ the corresponding fluctuation and $R_{(4)}$ is the rank-four Riemann curvature tensor $R_{iklm} \in \otimes^4 \dot T$, viz., in the cotangent bundle $\dot T$ of a given manifold $(M,g)$.

If we rescale this relationship down to the Planck scale $L_p$, by defining the light crossing time as $\tau = L$ and the Planck Time $\tau_p$, the Einstein equations retain their validity down to the Planck scale, even if metric fluctuations over a scale larger than $L_p$ can occur, extending the approach used in the Minkowski spacetime, making the CA structure more general and based on the properties of the spacetime.
We find that these fluctuations can give rise to a relationship between the curvature tensor and spacetime fluctuations that holds down to the Planck scales.
Once is fixed a characteristic spatial or temporal scale, $L$ or $\tau$, like in the building of a lattice structure, it corresponds to the introduction of fluctuations of the averaged quantity over $L^3$ of the proper energy $\bar E$. If we set $\bar E = \Delta E$ and $\tau = \Delta t$, considering fluctuations as large as the energy and time values considered, we can write a Heisenberg relationship that involves the Riemann tensor and the proper time,
\begin{equation}
 \Delta E\times \Delta t=\hbar \left(\frac{\tau}{\tau_p}\right)^2\frac{L^2}{g^2}R_{(4)}
 \label{ind1}
\end{equation}
equivalently one can write $\tau/\tau_p = L/L_P$.
At Planck scales Eq. \ref{ind1} is written as
\begin{equation}
 \Delta E \times \Delta t=\hbar \frac{L_p^2}{g^2}R_{(4)}=\hbar \left( \Delta \left(\frac{\Delta g}{g}\right) + \left(\frac{\Delta g}{g}\right)^2\right)
\label{ind2}
\end{equation}
where $\Delta E$ is averaged on the volume $L^3$ of a 3D space-like hypersurface $\sigma$ here considered, preserving the continuity of Einstein's equations down to the Planck scale, including the equivalence between Einstein-Rosen bridges and Einstein-Podolsky-Rosen states (ER=EPR) and graviton exchanges, as described in Ref. \cite{erepr}.

The indetermination relationship involving Planck's scales, here discussed, thus becomes an equivalence in a Minkowski-like manifold and a lattice structure. Quantum indetermination can arise from the lattice-like effects of spacetime fluctuations applied to deterministic continuous systems down to the Planck scales, as occurs to Einstein's equations, for which continuity holds. This is of course compatible with the holographic principle where any cell occupies a volume $L \times L^2_P$ and any spatial region with magnitude $L$ cannot contain more than $L^3/(L~L^2_P)=L^2/L^2_P$ cells; this agrees with the holographic principle for which the maximum of bit numbers stored in a region with characteristic length $L$ is at all effects $L^2/L^2_P=\tau/\tau_p $, in agreement with Eq. \ref{ind1} that can rewritten as 
\begin{equation} 
\Delta E \times \Delta t= \hbar~\frac{L^2}{g^2} R_{(4)}N_{bM} 
\end{equation}
When one indicates $N_{bM}=L^2/L^2_P$ the maximum number of bits there stored. QM appear as emerging from a lattice structure with the Holographic Principle (see e.g. \cite{licata2}).

This information can be the main core of an interpretation of the physics of CA in the periodic deterministic systems we will discuss below or be stored as a particle, according to the Hamiltonian of the system considered such as the Standard Model in a lattice system \cite{preparata,preparata1,preparata2}, which can provide the characteristic levels of the energy exchanges, interactions and time intervals of its quanta.

\subsection{Periodic deterministic systems}
When one considers a periodic model, it holds an  \(SU(2)\) symmetry, related to the rotation group \cite{thooft}, which is a subgroup of the Poincar\'e group. 
The elementary building blocks here considered consist on a CA system that updates itself at every time step, of duration \(\d t\), and then, after a period\  \(T=N\,\d t\), it returns to its initial position. They behave like gears that, cyclically rotating, concur to generate the perceived randomness of quantum mechanics when the dynamics has support in a lattice, where the time coordinate of the manifold is divided in discrete intervals $\tau$ and can be extended up to the hypothesis of a countably infinite lattice where the Hamiltonian eigenvalues will result in any case periodic \cite{thooft}.
Each single element of this construction with a finite number of states can thus be assumed to be periodic in time with a $SU(2)$ symmetry in a discrete time-quantized manifold. When one extends this procedure to the continuum, the Hamiltonian would need to be linearly dependent from the linear momentum like in the $H=xp$-class of dynamical systems already discussed.

Deterministic models can be seen as consisting of elementary cells inside which the data just oscillate in periodic orbits with $SU(2)$ symmetry. Rotation means angular momentum, as the invariant in the Poincar\'e group corresponding to rotation is the angular momentum. The energy eigenstates can be interpreted as 
the eigenstates $| m \rangle$ of L$_3$, the so-called $z-$component in a three dimensional rotator. The distribution of the eigenstates of these cells behave as being produced by the infinite-components Majorana equation (Majorana Tower), generated by the group of Lorentz boosts belonging to the Poincar\'e group of spacetime transformations. Of course, finite groups of rotators correspond to finite subgroups of the Majorana Tower, where 
the matrix elements ${}^x\langle r |s\rangle^p\) can be deduced from recursion relations, by fixing in this simple example where $H=\omega n$ and $x=s/\sqrt{\ell}$, $p=s/\sqrt{\ell}$,
such as
\begin{eqnarray} 
2r{\,}^x\langle r |s\rangle^p=
\langle \ell,s | a_x - i a_y | \ell, s+1\rangle 
\,{\,}^x\langle r | s-1\rangle^p \nonumber
\\
- 2 \langle \ell,s | b_x + i b_y | \ell-1, s-1 \rangle
\,{\,}^x\langle r |s+1\rangle^p\ ,
\qquad 
\label{recursion}
\end{eqnarray}
in combination with \({\,}^x\langle r|s\rangle^p={\,}^p\langle s|r\rangle^{x\,*}\) and more with the cyclic relations from Ref  \cite{thooft} and \cite{Majorana:NC:1932}, involving the infinitesimal Lorentz transformations in the variables $(ct,x,y,z)$,
\begin{eqnarray}
&&a_x=i \left(\begin{array}{cccc}0 & 0 & 0 & 0 \\0 & 0 & 0 & 0 \\0 & 0 & 0 & -1 \\0 & 0 & 1 & 0\end{array}\right); \, \,
a_y= i \left(\begin{array}{cccc}0 & 0 & 0 & 0 \\0 & 0 & 0 & 1 \\0 & 0 & 0 & 0 \\0 & -1 & 0 & 0\end{array}\right) \nonumber
\\
&&a_z=i \left(\begin{array}{cccc}0 & 0 & 0 & 0 \\0 & 0 & -1 & 0 \\0 & 1 & 0 & 0 \\0 & 0 & 0 & 0\end{array}\right);,
\label{ia}
\end{eqnarray}
and
\begin{eqnarray}
&&b_x= -i \left(\begin{array}{cccc}0 & 1 & 0 & 0 \\1 & 0 & 0 & 0 \\0 & 0 & 0 & 0 \\0 & 0 & 0 & 0\end{array}\right)
b_y= -i \left(\begin{array}{cccc}0 & 0 & 1 & 0 \\0 & 0 & 0 & 0 \\1 & 0 & 0 & 0 \\0 & 0 & 0 & 0\end{array}\right);  \nonumber 
\\
&&b_z= -i \left(\begin{array}{cccc}0 & 0 & 0 & 1 \\0 & 0 & 0 & 0 \\0 & 0 & 0 & 0 \\0 & 1 & 0 & 0\end{array}\right).
\label{ib}
\end{eqnarray}
\\
obtaining a Majorana equation that relates the coefficients of the CA with the infinitesimal Lorentz transformations of Eq \ref{ia} and \ref{ib} where the energy $E$ depends on the angular momentum that characterizes the CA and is positive-defined,
\begin{equation}
\left[ W + (\alpha, p) - \frac{E}{\ell + \frac 12}\right] \Psi = 0.
\label{majorana}
\end{equation}
where $W$ is the general energy from the Hamiltonian, $\alpha$ the set of Dirac  matrices, $p$ the momentum, $\ell$ the angular momentum eigenvalue and $E$ the energy considered to build the lattice structure.
In the limit $\delta t \rightarrow 0$, with infinite period $T$ that corresponds to $\ell=0$ turns this system into a point moving continuously along a circle, which behaves just like the standard harmonic oscillator.
Any CA can be seen as an excited Majorana states of the state $\ell=0$.
It is easy to show that down to the Planck scales one finds from Eq. \ref{majorana} the rules dictated by the holographic principle for the energy $E$ and the information there stored. The larger is $\ell$ the smaller is the energy and information density contained in a 3D hypervolume. To preserve the total information content integrated in the hypervolume, the latter has to grow linearly with $\ell$, with their corresponding vacuum and antivacuum states that grow with together with their entropy.
As previously said, depending on the Hamiltonian, these excited states of vacuum described by the sets of CA, can assume mass breaking for these states the infinite tower ruling the CA reducing it to the value of $\ell=1/2$ if there exists the Majorana neutrino.

\section{Conclusions}
We formulate a Majorana representation of the cellular automata for the ontological formulation of quantum mechanics and give the interpretation of the CA in terms of symmetries of spacetime.
From the Poincar\'e group of spacetime transformations and the subgroups of the Lorentz transformations and spatial rotations we obtain a relationship in terms of a Majorana-Dirac equation that gives the eigenvalues for the coefficients of the vector states describing the basic dynamics in a lattice-like structure that can take its origins from the properties of spacetime at Planck's scales. 
This represents a deep link between the basic fabric constituents of spacetime represented by the transformation groups and corresponding invariants and the structure of cellular automata that represent the fundamental building blocks of OQM.
The indetermination relationship obtained from Einstein's equations shows that the scale at which determinism can become or remain manifest is the Planck scale, where OQM interpretation can be obtained form a QM system equivalent to a deterministic dynamical system, supporting also a new interpretation of nonlocality in the ER=EPR scenario, making a parallelism between a deterministic Einstein Rosen bridge and entangled EPR states \cite{erepr}, where one joins elementary cells into a construction where they interact, again allowing only deterministic interaction laws mathematically closely related to the search of prime numbers through the P\'olya-Hilbert approach to the Riemann Hypothesis \cite{riemannft}.
In other words, what is normally thought of as being classical stochastic quantum mechanics it can be attributed to the effect of fast, almost hidden, variables that in any case support Bell's inequalities down to the Planck scale, directly from the texture of energy and spacetime fluctuations. 
The concept of ontological QM is related to dynamical systems and variables rapidly oscillating at the Planck scale - where we are forced to revise the ordinary continuous concepts of space and time - something that is at all effects similar to a set of hidden variables and that forms particles. Ontological is intended therefore as a global reflection on the languages of physics, classical and quantum, to set the conceptual conditions for their unification.
In this view, the Planck scale becomes a necessary scale where a lattice-like structure naturally arise and where one can find a whole topography of the ``non-local'', both below and above the Planck scale. Therefore, in a future and more complete formulation of these phenomena, QM, quantum field theory and GR will have to converge to a common language and set of concepts.

\end{document}